\documentclass{PoS}
\usepackage{graphicx}
\usepackage{epstopdf}
\usepackage{epsfig}
\usepackage{graphics}
\usepackage{amssymb}
\usepackage{amsmath}

\title{Jet splitting functions in the vacuum and in a QCD medium}

\ShortTitle{Jet splitting functions}

\author{%\speaker{Hai Tao Li}\\
        Hai Tao Li \\
        Los Alamos National Laboratory, Theoretical Division, Los Alamos, NM, 87545, USA\\
        E-mail: \email{haitaoli@lanl.gov}}

\author{Ivan Vitev\\
        Los Alamos National Laboratory, Theoretical Division, Los Alamos, NM, 87545, USA\\
        E-mail: \email{ivitev@lanl.gov}}

\abstract{The two-prong structure related to the leading subjets inside a reconstructed jet opens new avenues toward 
precision constraints on the in-medium modification of parton showers. 
In this talk, we present the first resummed calculation of the soft-dropped groomed momentum sharing distribution, 
or the jet splitting function, in heavy-ion collisions for both  light jets and heavy flavor tagged jets. 
Existing light jet splitting function data from the STAR experiment at RHIC and the CMS experiment at LHC 
can be understood in the unified framework of soft-collinear effective theory with Glauber gluon interactions. 
For heavy flavor jets, very interestingly,  the momentum sharing distribution of b-tagged jets is more strongly 
modified in comparison to the one for light jets, which provides a novel handle on mass corrections to in-medium parton showers that are at present difficult to constraint using inclusive heavy meson production. 
 }

\FullConference{International Conference on Hard and Electromagnetic Probes of High-Energy Nuclear Collisions\\
        30 September - 5 October 2018\\
        Aix-Les-Bains, Savoie, France}

\begin{document}

\section{Introduction}

Understanding the production and substructure of hadronic light jets and heavy flavor jets  
is crucial to test perturbative Quantum Chromodynamics (QCD). 
In heavy-ion collisions,  jet related observables have been widely 
used to study the fundamental thermodynamic and transport properties of a new deconfined state of matter, the quark-gluon plasma (QGP). 
In general, jet substructure, for a review paper see~\cite{Larkoski:2017jix}, 
is more sensitive to parton mass effects due to the smaller intrinsic scales involved, when compared to the inclusive jet production. 

In this talk, we present the first resummed predictions for the momentum sharing distribution, or jet splitting function, for  light jets and heavy flavor tagged jets in heavy-ion collisions~\cite{Li:2017wwc}, which is dominated by the first hard branching during the jet evolution. The momentum sharing distribution is constructed  using the soft-drop grooming algorithm~\cite{Larkoski:2014wba}. The momentum sharing variable and the grooming constraint are defined as 
\begin{align}
    z_g=\frac{\min(p_{T1}, p_{T2})}{p_{T1}+p_{T2}}~, \quad  z_g>z_{\rm cut} \left(\frac{\Delta R_{12}}{R} \right)^\beta \;, 
    \label{zg}
\end{align}
where $p_{T1}$ and ${p_{T2}}$ are the transverse momenta for the subjets. 
The variables $\Delta R_{12}$ and $R$ are the distance between two subjets and  the radius of the original jet.  
In the limit of large jet energies, the distribution of $z_g$ maps directly onto the widely used  Altarelli-Parisi splitting functions.

\section{Jet splitting functions }
The jet splitting functions in the vacuum can be derived from  the double differential branching distribution for parton $j$  that initiates the  jet. For light jets we have 
\begin{align} \label{eq:split}
    \left(\frac{dN^{\rm vac}}{dz_g d\theta_g}\right)_{j}  = \frac{\alpha_s}{\pi} \frac{1}{\theta_g} \sum_{i} P_{j\to i \bar{i}}^{\rm vac}(z_g)~,
\end{align}
where $\theta_g$ is defined as $\Delta R_{12}/R$ and the sum runs over all possible channels. 
When  $\theta_g\sim 0$, the fixed order prediction is not valid due to the collinear divergence and  resummation 
is necessary, which was performed up to the modified leading-logarithmic (MLL) accuracy  in Ref.~\cite{Larkoski:2014wba}. 
The resummed distribution for the jets initiated by a massless quark or a gluon  is 
\begin{align} \label{eq:mll}
    \frac{dN_j^{\rm vac,MLL}}{ dz_g d\theta_g} =& \sum_{i} \left(\frac{dN^{\rm vac}}{dz_g d\theta_g}\right)_{j\to i \bar{i}} 
     \underbrace{\exp \left[-\int_{\theta_g}^1 d\theta \int_{z_{\rm cut}}^{1/2} dz  \sum_{i} \left(\frac{dN^{\rm vac}}{dz d\theta}\right)_{j\to i \bar{i}}  \right]}_{\rm Sudakov~Factor}~.
\end{align}
The normalized joint probability distribution of subjets then reads
\begin{align} \label{eq:mll2}
    p(\theta_g,z_g)\big|_{j}=\frac{dN_j^{\rm vac,MLL}}{ dz_g d\theta_g} \bigg/\left(\int_{0}^1 d\theta \int_{z_{\rm cut}}^{1/2} dz  \frac{dN_j^{\rm vac,MLL}}{ dz d\theta} \right)~. 
\end{align}
For heavy flavor jets we need to re-derive Eqs.~(\ref{eq:split}, \ref{eq:mll} and \ref{eq:mll2})  
by replacing the light parton splitting kernel with the corresponding  heavy flavor ones, see Ref.~\cite{Li:2017wwc}.  
The final probability distribution is defined as the convolution of the normalized hard part and the jet  joint probability distribution. 

In the presence of a QCD medium, it was demonstrated that the vacuum distributions must be replaced by the ones with medium corrections 
for each possible partonic channel
\begin{align}
    \left(\frac{dN^{\rm full}}{dz d\theta_g}\right)_{j} = \left(\frac{dN^{\rm vac}}{dz  d\theta_g}\right)_{j}  
    + \left(\frac{dN^{\rm med}}{dz  d\theta_g }\right)_{j} \; .      
\end{align}
The  full-set of medium-induced splitting kernels have 
been calculated in the framework of $\rm SCET_{\rm (M,) G}$~\cite{Ovanesyan:2011kn,Kang:2016ofv} 
and recently applied to describe inclusive heavy flavor jet quenching~\cite{Li:2018xuv}.

\section{Numerical results}

The modification of the jet splitting functions in heavy-ion collisions is defined as the ratio of the $z_g$ distributions in the medium and the vacuum. The uncertainties are obtained by varying the coupling g between the jet and the QGP as indicated in each figure and all the scales in the formalism. 

\begin{figure}[h]
    \centering
    \includegraphics[scale=0.6]{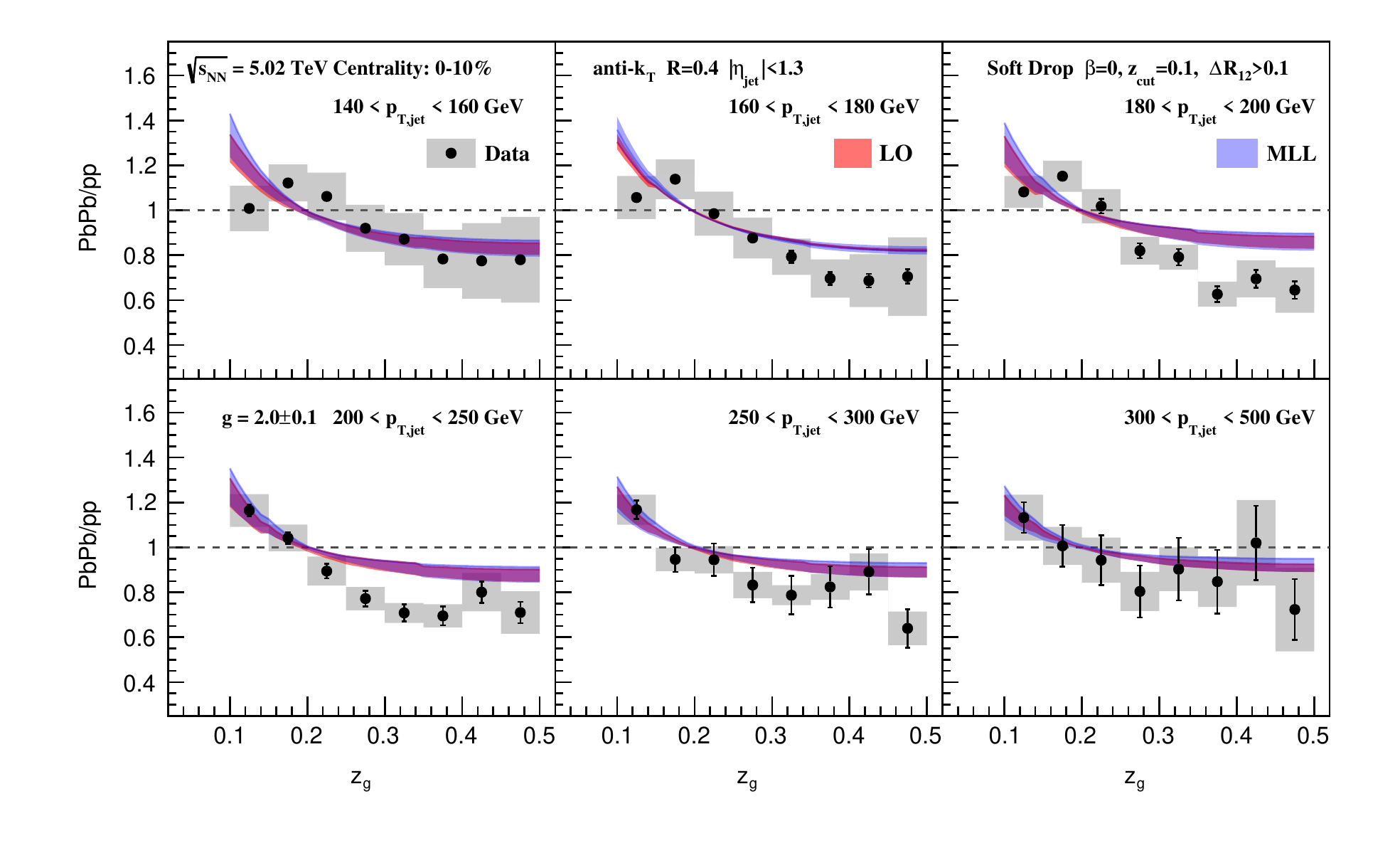}
    \vspace{-0.8cm}
    \caption{Comparison of theoretical predictions for the in-medium $z_g$ distribution modification to CMS measurements in Pb+Pb collisions. }
    \label{fig:cms}
\end{figure}

In Fig.~\ref{fig:cms} we present the modification of the resummed momentum sharing distributions  of 
inclusive light jets over different kinematic ranges in 0-10\% central Pb+Pb collisions  at $\sqrt{s_{\rm NN}}=5.02$~TeV 
compared to  measurements by the CMS collaboration at the LHC Run II~\cite{Sirunyan:2017bsd}.  
An additional cut on the distance between the two subjets $\Delta R_{12}>0.1$ is applied due  to the detector resolution, 
which makes it possible to provide a LO prediction as well~\cite{Chien:2016led}. For the same reason, we notice that there is little difference  
between the fixed order  and resummed  calculations, and theoretical predictions agree  well with the experimental measurements.  

\begin{figure}
    \centering
    \includegraphics[scale=0.3]{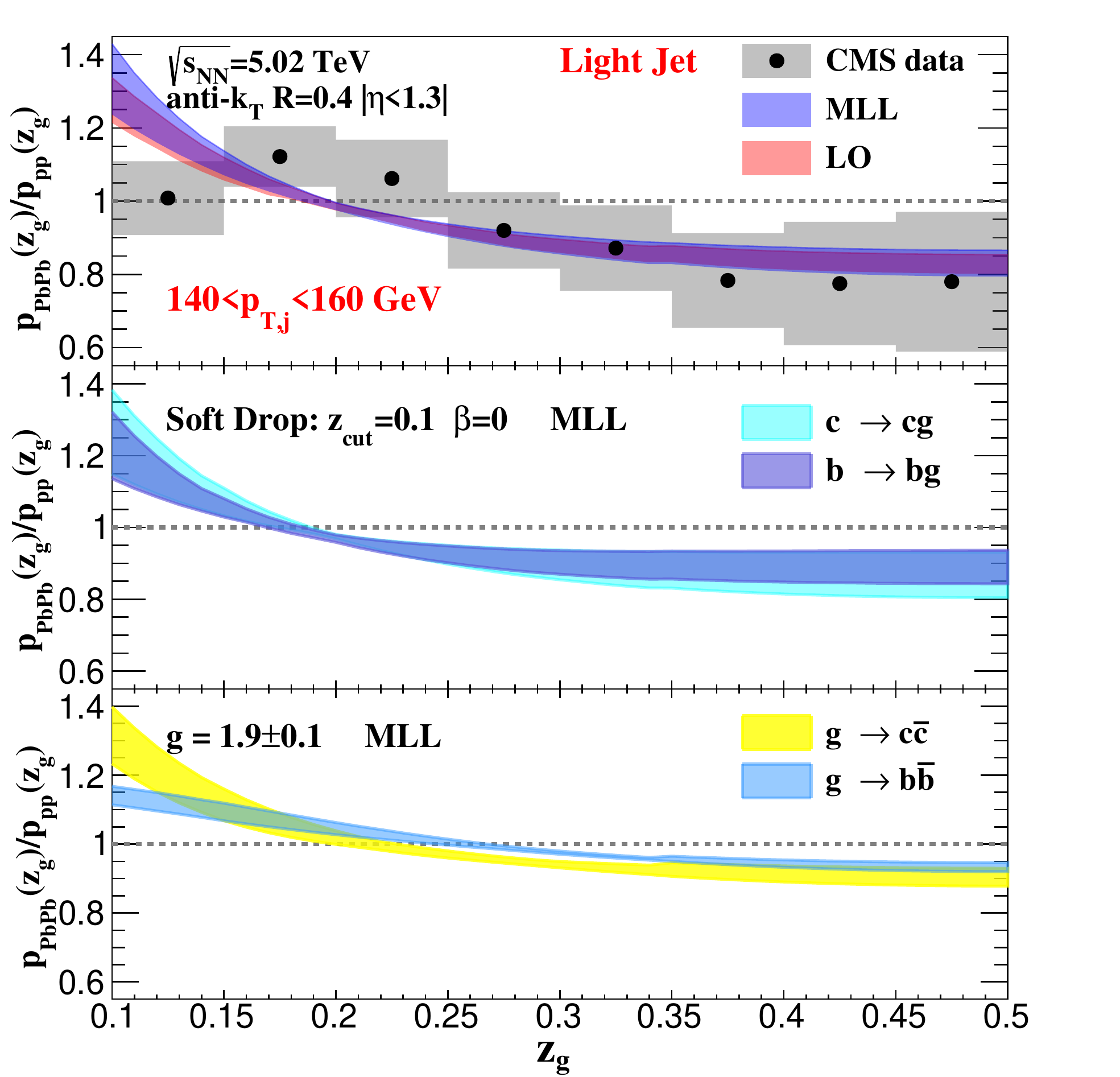} \hspace*{1cm}
        \includegraphics[scale=0.3]{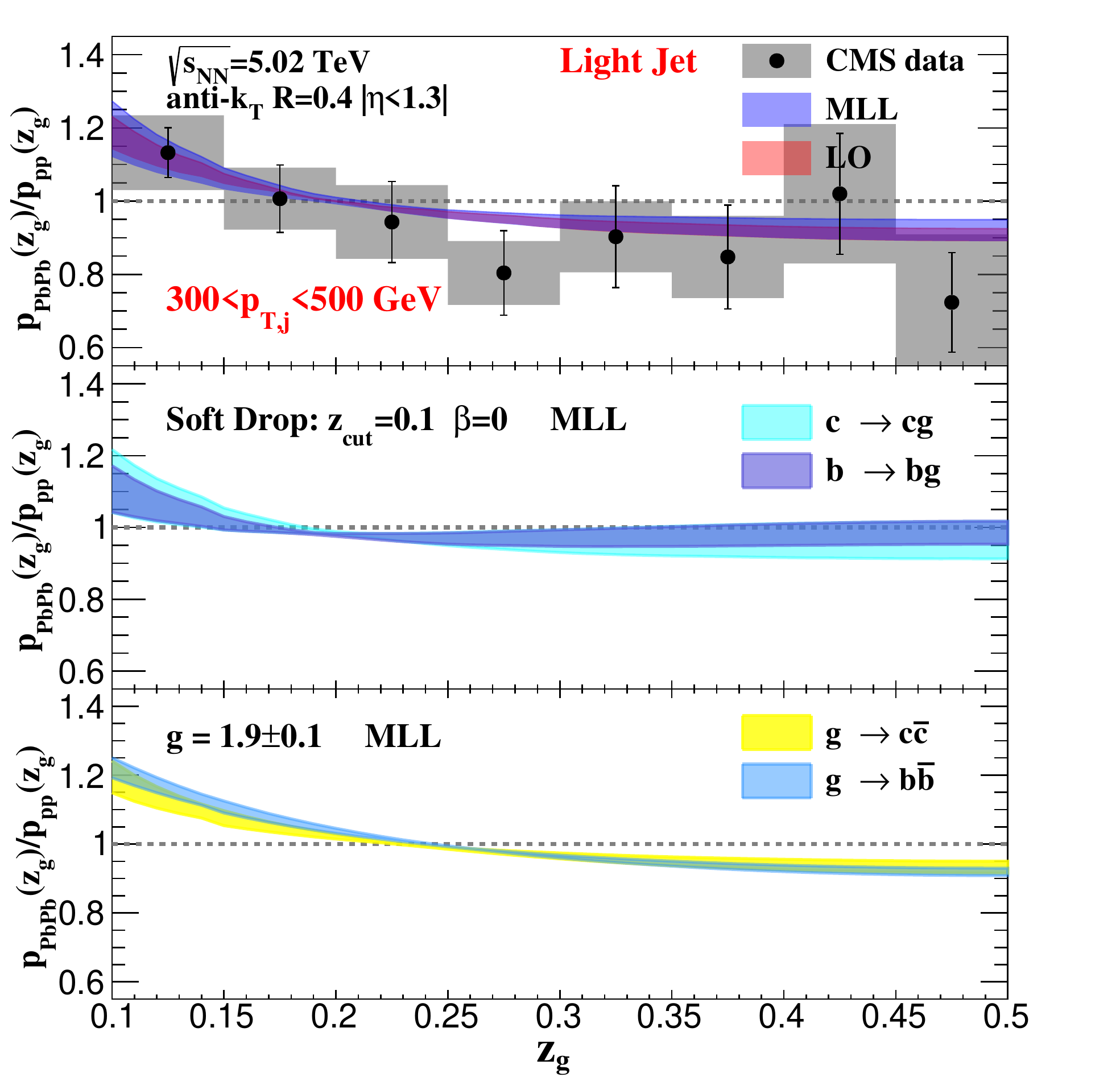}
            \vspace{-0.2cm}
    \caption{
    Comparison of the QGP-induced modifications for the light jets (upper) and heavy flavor tagged jets (middle and lower) splitting functions
    in 0-10\% central Pb+Pb collisions at $\sqrt{s_{\rm NN}}$= 5.02 TeV for two $p_T$ bins 140< $p_{T,j}$<160 GeV (left) and 300< $p_{T,j}$<500 GeV (right).}
    \label{fig:cmsheavy}
\end{figure}

\begin{figure}
\begin{minipage}{16pc}
\includegraphics[scale=0.3]{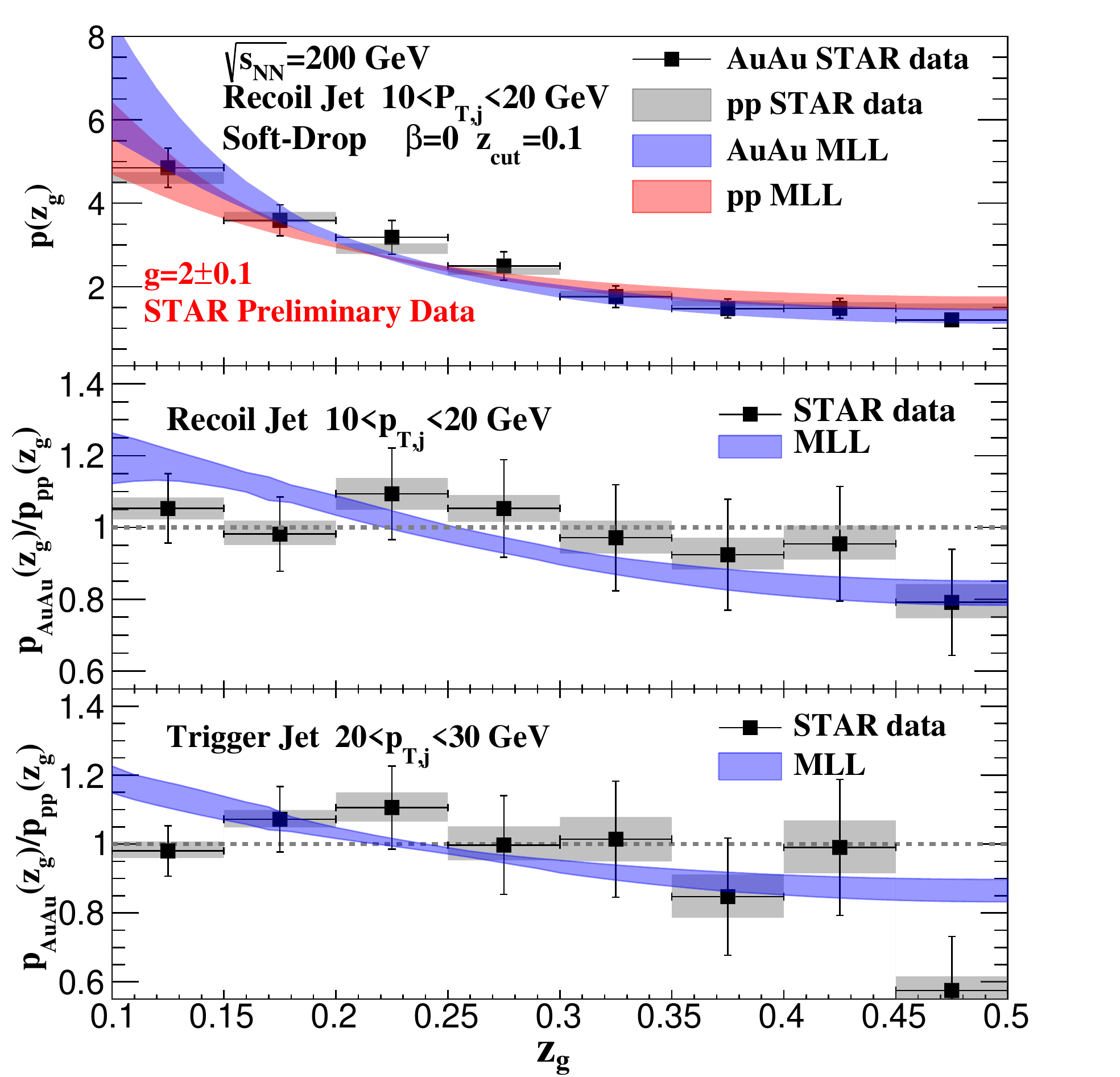}
\caption{\label{fig:ppstar}The distribution of $z_g$ and its modification for recoil and trigger light jets at $\sqrt{s_{\rm NN}}$= 200 GeV Au+Au collisions. }
\end{minipage}\hspace{2pc}%
\begin{minipage}{16pc}
 \vspace{-0.5cm}
 \includegraphics[scale=0.35]{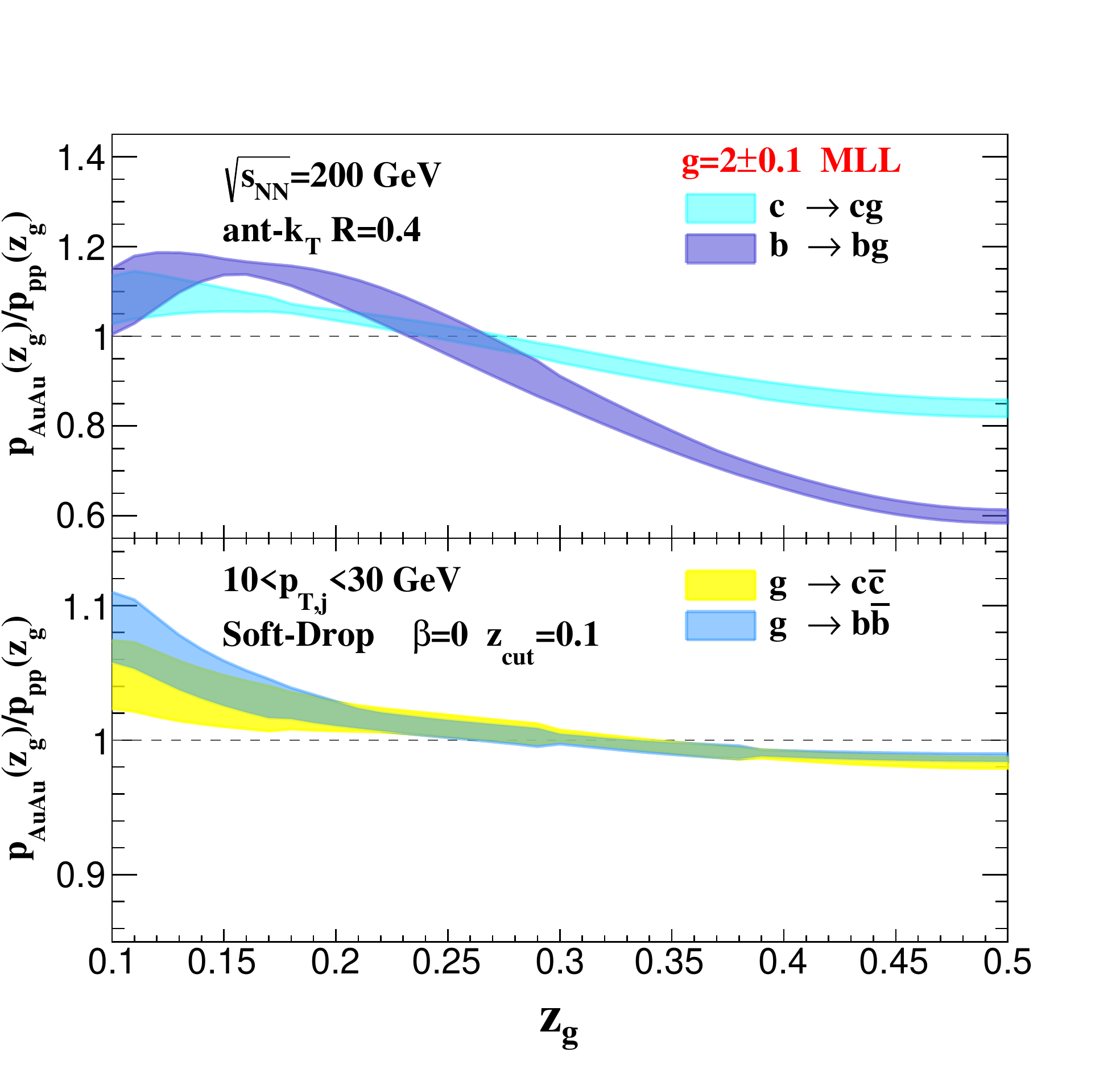}
 \vspace{-1cm}
\caption{\label{fig:bjet} The modifications of the splitting functions for heavy flavor tagged jets at $\sqrt{s_{\rm NN}}$= 200 GeV Au+Au collisions.  }
\end{minipage} 
\end{figure}
%\vspace{-1cm}

In Fig.~\ref{fig:cmsheavy} the first predictions for the modification of prompt $b$-jet and $c$-jet substructure  
are given in the middle panel and the modification of $g\to Q\bar{Q}$ jet substructure
is shown in the lower panel. We select jets in the kinematic regions $140<p_{T}<160$~GeV (left) and $300<p_{T}<500$~GeV (right). We find that jet quenching effects for $p(z_g)$ are comparable to those of light jets and we expect it can be measured by the LHC experiments. The mass effect slowly vanishes with increasing jet energy.

Figure~\ref{fig:ppstar} shows the light jet momentum sharing distribution and its modification in Au+Au collisions at $\sqrt{s_{\rm NN}}=200$~GeV,
where our predictions are compared to the measurements by the STAR collaboration~\cite{Kauder:2017cvz} for the trigger and recoil jets. 
Both  the  splitting  functions  and  their modification  are in good agreement with data after we consider the  experimental uncertainties.  As we can see, the MLL results and the measured modifications are smaller than those at the LHC.

Before we move on to the heavy flavor jets with lower transverse momenta, 
let us introduce the analytic first principles expectations for the jet splitting function modification at the leading order.  
We focus on the $z$ dependence of parton branching in the vacuum and medium-induced splitting kernels, noticing that in the medium one has two mass dependent propagators~\cite{Kang:2016ofv}.  If we write the $z_g$ distribution in heavy-ion collisions as the normalized  $(p_{\rm pp} + p_{\rm med})$,  when $z_g$ is small we can predict the following jet splitting function modifications in order of decreasing strength
\begin{equation}
\frac{p_{med}^{Q\rightarrow Qg}(z_g)}{p_{\rm pp}^{Q\rightarrow Qg}(z_g)} \sim \frac{1}{z_g^2},  \ \  \; 
\frac{p_{med}^{j\rightarrow i\bar{i}}(z_g)}{p_{\rm pp}^{j\rightarrow i\bar{i}}(z_g)} \sim \frac{1}{z_g},   \ \ \; 
\frac{p_{med}^{g\rightarrow Q\bar{Q}}(z_g)}{p_{\rm pp}^{g\rightarrow Q\bar{Q}}(z_g)} \sim {\rm const} .
\label{order}
\end{equation}
Numerical results for the heavy flavor momentum sharing distribution ratios  in  Au+Au to p+p collisions  
at $\sqrt{s_{\rm NN}}=200$ GeV are presented in Fig.~\ref{fig:bjet}.  
We consider the jets with $10<p_{T, j}<30$~GeV where our analysis suggests that mass effects on parton shower formation are large for b-jets.
The $b\to bg$ channel  exhibits  much larger in-medium effects, while the $c\to cg$ channel evolving in QCD medium is similar to light jets. 
This unique inversion of the mass hierarchy of jet quenching effects is in perfect agreement  with Eq.~(\ref{order}),
and so is the very small nuclear modification for  the $g\to Q\bar{Q}$ channel in the bottom panel of Fig.~\ref{fig:bjet}.

\section{Summary}

The first resummed predictions of the soft-drop groomed momentum sharing distributions in heavy-ion collisions were presented  based on the framework of recently developed soft-collinear effective theory with Glauber gluon interactions and quark mass effects. The theoretical calculations for light jets agree well with the recent measurements in Au+Au and Pb+Pb reactions over a wide range of center-of-mass energies which validates the theoretical approach.  Most importantly we found that in the kinematic regime where parton mass plays the most important role the momentum sharing distribution for prompt b-jets is the largest, which is  driven by the heavy quark mass and is measurable at the LHC and by the future sPHENIX experiment at the RHIC.

\subsection*{Acknowledgments}
This work is supported on US Department of Energy, Office of Science under Contract No. DE-AC52-06NA25396, the DOE Early Career Program, and the LDRD program at LANL.


\begin{thebibliography}{99}

%\cite{Larkoski:2017jix}
\bibitem{Larkoski:2017jix} 
  A.~J.~Larkoski, I.~Moult and B.~Nachman,
%  ``Jet Substructure at the Large Hadron Collider: A Review of Recent Advances in Theory and Machine Learning,''
  arXiv:1709.04464 [hep-ph].
  %%CITATION = ARXIV:1709.04464;%%
  %70 citations counted in INSPIRE as of 26 Nov 2018



%\cite{Li:2017wwc}
\bibitem{Li:2017wwc} 
  H.~T.~Li and I.~Vitev,
 % ``Inverting the mass hierarchy of jet quenching effects with prompt $b$-jet substructure,''
  arXiv:1801.00008 [hep-ph].
  %%CITATION = ARXIV:1801.00008;%%
  %8 citations counted in INSPIRE as of 26 Nov 2018



%\cite{Larkoski:2014wba}
\bibitem{Larkoski:2014wba} 
  A.~J.~Larkoski, S.~Marzani, G.~Soyez and J.~Thaler,
%  ``Soft Drop,''
  JHEP {\bf 1405}, 146 (2014)
%  doi:10.1007/JHEP05(2014)146
  [arXiv:1402.2657 [hep-ph]].
  %%CITATION = doi:10.1007/JHEP05(2014)146;%%
  %298 citations counted in INSPIRE as of 26 Nov 2018



%\cite{Ovanesyan:2011kn}
\bibitem{Ovanesyan:2011kn} 
  G.~Ovanesyan and I.~Vitev,
%  ``Medium-induced parton splitting kernels from Soft Collinear Effective Theory with Glauber gluons,''
  Phys.\ Lett.\ B {\bf 706}, 371 (2012)
%  doi:10.1016/j.physletb.2011.11.040
  [arXiv:1109.5619 [hep-ph]].
  %%CITATION = doi:10.1016/j.physletb.2011.11.040;%%
  %57 citations counted in INSPIRE as of 26 Nov 2018



%\cite{Kang:2016ofv}
\bibitem{Kang:2016ofv} 
  Z.~B.~Kang, F.~Ringer and I.~Vitev,
%  ``Effective field theory approach to open heavy flavor production in heavy-ion collisions,''
  JHEP {\bf 1703}, 146 (2017)
%  doi:10.1007/JHEP03(2017)146
  [arXiv:1610.02043 [hep-ph]].
  %%CITATION = doi:10.1007/JHEP03(2017)146;%%
  %26 citations counted in INSPIRE as of 26 Nov 2018

%\cite{Li:2018xuv}
\bibitem{Li:2018xuv} 
  H.~T.~Li and I.~Vitev,
  %``Inclusive heavy flavor jet production with semi-inclusive jet functions: from proton to heavy-ion collisions,''
  arXiv:1811.07905 [hep-ph].
  %%CITATION = ARXIV:1811.07905;%%



%\cite{Sirunyan:2017bsd}
\bibitem{Sirunyan:2017bsd} 
  A.~M.~Sirunyan {\it et al.} [CMS Collaboration],
 % ``Measurement of the Splitting Function in $pp$ and Pb-Pb Collisions at $\sqrt{s_{_{\mathrm{NN}}}} =$ 5.02 TeV,''
  Phys.\ Rev.\ Lett.\  {\bf 120}, no. 14, 142302 (2018)
%  doi:10.1103/PhysRevLett.120.142302
  [arXiv:1708.09429 [nucl-ex]].
  %%CITATION = doi:10.1103/PhysRevLett.120.142302;%%
  %32 citations counted in INSPIRE as of 26 Nov 2018


%\cite{Chien:2016led}
\bibitem{Chien:2016led} 
  Y.~T.~Chien and I.~Vitev,
%  ``Probing the Hardest Branching within Jets in Heavy-Ion Collisions,''
  Phys.\ Rev.\ Lett.\  {\bf 119}, no. 11, 112301 (2017)
%  doi:10.1103/PhysRevLett.119.112301
  [arXiv:1608.07283 [hep-ph]].
  %%CITATION = doi:10.1103/PhysRevLett.119.112301;%%
  %34 citations counted in INSPIRE as of 26 Nov 2018

%\cite{Kauder:2017cvz}
\bibitem{Kauder:2017cvz} 
  K.~Kauder [STAR Collaboration],
%  ``Measurement of the Shared Momentum Fraction $z_g$ using Jet Reconstruction in p+p and Au+Au Collisions with STAR,''
  Nucl.\ Part.\ Phys.\ Proc.\  {\bf 289-290}, 137 (2017)
%  doi:10.1016/j.nuclphysbps.2017.05.028
  [arXiv:1703.10933 [nucl-ex]].
  %%CITATION = doi:10.1016/j.nuclphysbps.2017.05.028;%%
  %9 citations counted in INSPIRE as of 26 Nov 2018




\end{thebibliography}
\end{document}